\RequirePackage{amsmath}

\documentclass[12pt]{iopart}
\pdfoutput=1
\usepackage{graphicx}
\usepackage{amsmath}
\usepackage{amsfonts}
\usepackage{mathtools}
\usepackage{amssymb}
\usepackage{xcolor}
\usepackage{subfig}

\colorlet{mylinkcolor}{blue!66!black!80}

\usepackage[colorlinks=true,linkcolor=mylinkcolor,citecolor=mylinkcolor,filecolor=cyan,urlcolor=mylinkcolor,breaklinks=true]{hyperref}

\usepackage[utf8]{inputenc}

\usepackage{newtxtext}
\usepackage{newtxmath}

\usepackage{cite}

\DeclareMathAlphabet{\mathcal}{OMS}{cmsy}{m}{n}

\begin{document}
\title{Cyclic Heat Engine with the Ising model: role of interactions and criticality}% Force line breaks with \\
\author{Gustavo A. L. For\~ao$^1$, Arya Datta$^2$, Carlos E. Fiore$^1$, and Andre C. Barato$^2$}
\address{
$^1$Instituto de F\'isica da Universidade de S\~ao Paulo, 05314-970 S\~ao Paulo, Brazil\\
$^2$Department of Physics, University of Houston, Houston, Texas 77204, USA}
%\eads{\href{mailto:hartich@theo2.physik.uni-stuttgart.de}{hartich@theo2.physik.uni-stuttgart.de}, \href{mailto:barato@theo2.physik.uni-stuttgart.de}{barato@theo2.physik.uni-stuttgart.de} and
%\href{mailto:useifert@theo2.physik.uni-stuttgart.de}{useifert@theo2.physik.uni-stuttgart.de}}
\date{\today}

\begin{abstract}

Heat engines that convert thermal energy into work are a cornerstone of classical thermodynamics and remain an active area of contemporary research. 
Notable examples include microscopic heat engines, trade-off relations between power and efficiency, and the attainability of Carnot efficiency at finite power. 
We propose a cyclic heat engine based on the Ising model, in which the thermodynamic cycle involves variations of both temperature and magnetic field. 
We analyze the one-dimensional and mean-field Ising models, which allow for simple analytical results and provide new insight into the role of interactions in cyclic heat engines.
In particular, we show that interactions can enhance both power and efficiency. Moreover, a system that does not operate as an engine in the absence of interactions can become an engine 
upon tuning the interaction strength. The mean-field model enables us to investigate the relevance of the phase transition for the performance of this Ising heat engine. Owing to the emergence 
of spontaneous magnetization, the mean-field model can still operate as an engine even when one of the magnetic fields is set to zero. Remarkably, when the work is maximized, 
we find that the optimal parameters are numerically consistent with this regime, in which one magnetic field vanishes and the cycle explores the phase transition.
We also consider an alternative cycle for the mean-field model, obtained by varying the interaction strength while keeping both temperatures below the critical temperature and setting the magnetic field to zero 
throughout the cycle. The power and efficiency of this cycle are analyzed as well. Finally, while our analytical results are valid for the limit of large period we use numerical simulations for finite periods and 
show that the power decreases monotonically with the period.

\end{abstract}
% \maketitle

%%%%%%%%%%%%%%%%%%%%%%%%%%%%%%%%%%%%%%%%%%%%%%%%%%%%%%%%%%%%%%%%%%%%%
\section{Introduction}
%%%%%%%%%%%%%%%%%%%%%%%%%%%%%%%%%%%%%%%%%%%%%%%%%%%%%%%%%%%%%%%%%%%%%

 Heat engines were a driving force of the Industrial Revolution. In theoretical physics, they also played a central role in the development of the formal framework of thermodynamics, which provided deep insight into the fundamental limitations of heat engines. Most prominently, the efficiency of heat engines is bounded by the universal Carnot efficiency, which depends only on the temperatures of the hot and cold reservoirs. 
 Despite its success, standard thermodynamics has important limitations. While it provides universal bounds on the performance of cyclic heat engines, 
it does not offer the tools required to analyze the power of finite-time heat engines operating away from the quasi-static regime.
Moreover, standard thermodynamics is restricted to macroscopic systems with negligible fluctuations. This assumption does not hold for more 
 recent experimental advances that allow for the manipulation and design of systems at the molecular and atomic scales.

Beyond standard thermodynamics, heat engines remain an active area of research. 
Early studies of finite-time heat engines, including endoreversible cycles, analyzed their performance within linear response or near-equilibrium frameworks \cite{curz75,band82,andr84,leff87},
with particular emphasis on the trade-off between power 
and efficiency. The modern framework of stochastic thermodynamics \cite{seif12,peli21,shir23,seif25} enables the investigation of heat engines beyond the linear response regime.

Several topics have been studied more recently within this framework. The efficiency at maximum power of finite-time heat engines, as well as the degree of 
universality of the Curzon–Ahlborn efficiency introduced in \cite{curz75}, were analyzed in  \cite{vdb05,gome06,schm08a,espo09,espo09b}. 
The realization of microscopic cyclic heat engines based on a single colloidal particle
has been explored both theoretically \cite{schm08b,espo10,izum11,tu14,bran15,raz16,ray17} and experimentally \cite{stee11,blic12,mart15,mart16,ross16,mart17}. 
Another line of research concerns active heat engines, in which the reservoir contains dissipative degrees of freedom such as bacteria. 
Their unusual behavior was first observed experimentally  \cite{kris16} and subsequently investigated theoretically \cite{zaki17,saha19,piet19,ekeh20,holu20,holu20b,kuma20,lee20,fodo21,gron21,datt22,alba23,wies24}. 
Engines that achieve Carnot efficiency at finite power output \cite{alla13,camp16,pole17,lee17,abiu20,lian25} have been proposed. 
Universal trade-off relations between power and efficiency have been derived, both for cyclic heat engines driven by an external periodic 
protocol \cite{shir16,koyu19} and for steady-state engines driven by constant thermodynamic forces \cite{piet18}. In the large field of information thermodynamics \cite{parr15},  
engines driven by information have been proposed \cite{toya10a,abre11,mand12,stra13,kosk14,ribe19,Pane22,malg22,saha23}.

Many of the aforementioned papers in stochastic thermodynamics analyze small systems made of a single constituent. 
However, heat engines made of many-body systems have also been considered \cite{golu13,vroy19,filh23,fora25}. 
In particular, a cyclic heat engine that operates at the critical point and comes arbitrarily close to Carnot efficiency at finite power 
has been proposed \cite{camp16}.  Nevertheless, the role of interactions and phase transitions  in a cyclic heat engines remains largely unexplored.

\begin{figure}%
    \hspace*{-2cm}%
    \includegraphics[height=0.85\textheight,angle=-90]{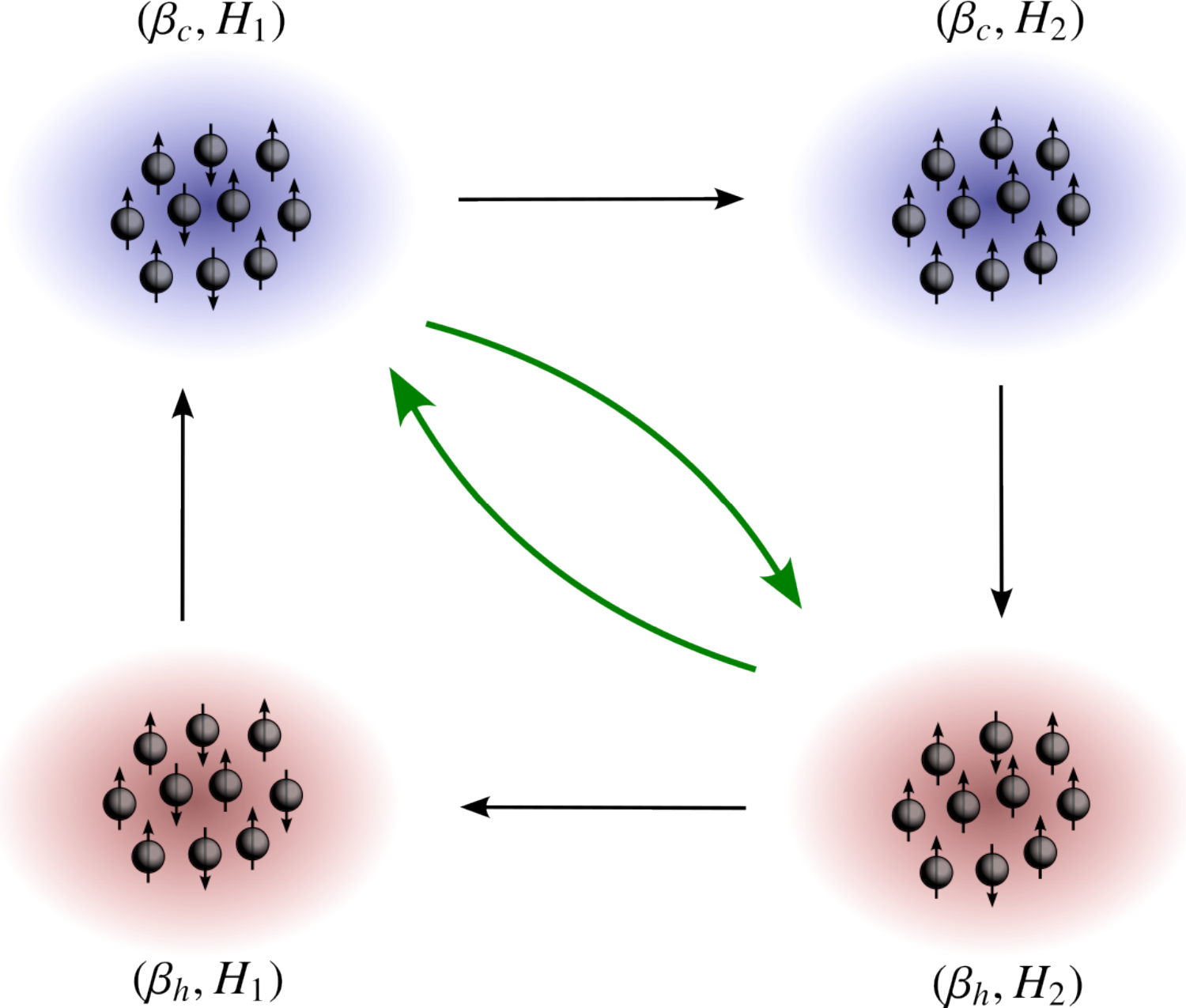}%
    \caption{Cyclic protocol for the Ising model. The thermodynamic parameters are the inverse temperature $\beta$ and the external magnetic field $H$. The changes in temperature are assumed to be instantaneous. Effectively, the system stays in the first part of the cycle for half of the period and in the third part of the cycle for the other half of the period, with the second and fourth parts having a negligible duration.}%
    \label{fig1}%
\end{figure}

Here we introduce a heat engine based on the paradigmatic Ising model. We consider the cycle illustrated in Fig. \ref{fig1}, 
in which both the external magnetic field and the temperature are varied. This model provides a natural framework to investigate
 the role of interactions and phase transitions in cyclic heat engines. In particular, we analyze how the interaction strength 
 affects work and efficiency. We study both the one-dimensional (1D) and the mean-field (MF) Ising models, which allow 
 for exact expressions for heat and work in the limit where the period is long compared to the relaxation times of the system. 
 While the 1D Ising model leads to simpler analytical expressions, the MF model, owing to the presence of a phase transition, 
 exhibits two distinct regimes that arise from spontaneous magnetization.

We now summarize our main results. Interactions enlarge the parameter region in which 
the system operates as a heat engine. In particular, for values of temperatures and external fields for which a noninteracting system does not extract work, the presence of interactions can render 
the system functional by enabling work extraction. For the 1D Ising model, this behavior is demonstrated analytically through a simple exact expression. 
Interactions can enhance both the extracted work and the efficiency: the work is maximized for ferromagnetic interactions, whereas the efficiency 
can be maximized for both ferromagnetic and antiferromagnetic interactions. In the MF model, the phase transition gives rise to two distinct operating regimes. 
In the first regime, the system functions as a heat engine even when one of the magnetic fields is set to zero, a behavior made possible by spontaneous magnetization. 
Remarkably, the maximum power is achieved precisely in this regime. In the second regime, the system operates as a heat engine with both magnetic fields set to zero, 
while the interaction strength is varied during the cycle. For this regime, we analyze the maximum work and the efficiency at maximum power.
In addition, we perform numerical simulations of the MF model to investigate the impact of a finite period on the power output. 
Our simulations show that the power decreases monotonically as the period increases. Finally, we note that the simplicity and analytical tractability of the 
1D and MF Ising models make our model a promising textbook example for illustrating the behavior of cyclic heat engines.

The paper is organized as follows. Sec. \ref{sec2} contains the definitions of the models and the general expressions for heat and work. 
Analytical results in the long-period limit are presented in Sec. \ref{sec3}.
Numerical results for finite periods are discussed in Sec. \ref{sec4}. Conclusions are given in Sec. \ref{sec5}.

%%%%%%%%%%%%%%%%%%%%%%%%%%%%%%%%%%%%%%%%%%%%%%%%%%%%%%%%%%%%%%%%%%%%%
\section{Model and exact expressions}
%%%%%%%%%%%%%%%%%%%%%%%%%%%%%%%%%%%%%%%%%%%%%%%%%%%%%%%%%%%%%%%%%%%%%
\label{sec2}

%%%%%%%%%%%%%%%%%%%%%%%%%%%%%%%%%%%%%%%%%%%%%%%%%%%%%%%%%%%%%%%%%%%%%
\subsection{Cyclic protocol}
%%%%%%%%%%%%%%%%%%%%%%%%%%%%%%%%%%%%%%%%%%%%%%%%%%%%%%%%%%%%%%%%%%%%%

The cycle shown in Fig.~\ref{fig1} consists of four parts. The total period of the cycle is denoted by $\tau$. 
The thermodynamic control parameters are the inverse temperature $\beta$ and the external magnetic field $H$. 
Throughout this paper, we use the inverse temperature $\beta = 1/(k_B T)$ instead of the temperature $T$, 
and we set Boltzmann's constant to $k_B = 1$.

During the first part of the cycle, the parameters are $H_1$ and $\beta_c$, and its duration is half of the period, $\tau/2$. 
At the end of this first part, the magnetic field is changed from $H_1$ to $H_2$. 
The second part of the cycle, corresponding to parameters $H_2$ and $\beta_c$, is assumed to have a negligible duration; 
during this step, the inverse temperature is instantaneously changed from $\beta_c$ to $\beta_h$. 
The third part of the cycle has parameters $H_2$ and $\beta_h$ and also lasts for a duration $\tau/2$. 
At the end of this third part, the magnetic field is switched back from $H_2$ to $H_1$. 
Finally, the fourth part of the cycle is again assumed to have negligible duration, during which the inverse temperature 
is instantaneously changed from $\beta_h$ back to $\beta_c$. Since both temperature changes are instantaneous, the cycle actually has only two parts, as represented in 
Fig.~\ref{fig1}.

We consider both the 1D and the MF Ising models undergoing this cycle. 
A key difference between these two variants is that the MF Ising model exhibits a phase transition, 
which has important implications for the performance of the heat engine. 
The Hamiltonian of the 1D Ising model is given by
\begin{equation}
E_\sigma = -J \sum_{i=1}^N \sigma_i \sigma_{i+1} - H \sum_{i=1}^N \sigma_i,
\label{eqE1d}
\end{equation}
where $J$ denotes the interaction strength, $N$ is the number of spins, and $\sigma_i = \pm 1$ 
represents the orientation of spin $i$. Periodic boundary conditions are assumed. 
For the MF Ising model, the Hamiltonian reads
\begin{equation}
E_\sigma = -\frac{J}{2N} \sum_{i=1}^N \sum_{j=1}^N \sigma_i \sigma_j - H \sum_{i=1}^N \sigma_i.
\label{eqEmf}
\end{equation}

Two quantities of central interest are the average magnetization per spin,
\begin{equation}
m_{\beta,H,J} \equiv \frac{1}{N} \sum_\sigma P_\sigma \sum_{i=1}^N \sigma_i,
\label{eqm}
\end{equation}
and the average energy per spin,
\begin{equation}
u_{\beta,H,J} \equiv \frac{1}{N} \sum_\sigma P_\sigma E_\sigma.
\label{equ}
\end{equation}
In both expressions, $P_\sigma \propto \exp(-\beta E_\sigma)$ denotes the Boltzmann distribution. 
The subscripts indicate the dependence of these quantities on the inverse temperature $\beta$, 
the magnetic field $H$, and the interaction strength $J$. 
We focus on the limit where the period $\tau$ is large compared to the relaxation times of the system. 
In this limit, exact expressions for heat and work can be obtained, and we employ the well-known 
expressions for the magnetization and average energy of the Ising model in the thermodynamic limit \cite{baxt82}.

For the 1D Ising model, the average magnetization per spin in the thermodynamic limit is given by
\begin{equation}
m_{\beta,H,J} = \frac{\sinh(\beta H)}{\sqrt{\sinh^2(\beta H) + e^{-4\beta J}}}.
\label{eqm1d}
\end{equation}
The corresponding average energy per spin reads
\begin{equation}
u_{\beta,H,J}
=
- H\, m_{\beta,H,J}
-
J\,
\frac{
\sinh^2(\beta H) + e^{-2\beta J}
}{
\cosh(\beta H)\sqrt{\sinh^2(\beta H)+e^{-4\beta J}}
+
e^{-2\beta J}
}.
\label{equ1d}
\end{equation}

For the MF Ising model, the average magnetization per spin is determined by the solution of the 
transcendental equation
\begin{equation}
m_{\beta,H,J} = \tanh(\beta J m_{\beta,H,J} + \beta H).
\label{eqmmf}
\end{equation}
In the presence of a phase transition leading to spontaneous magnetization, the magnetization 
at vanishing magnetic field, $H = 0$, is defined as
\begin{equation}
m_{\beta,J} \equiv m_{\beta,0,J}.
\label{eqm0}
\end{equation}
The average energy per spin in the MF model, expressed in terms of the solution of the 
transcendental equation, is given by
\begin{equation}
u_{\beta,H,J} = - H m_{\beta,H,J} - \frac{J}{2} \left(m_{\beta,H,J}\right)^2 .
\label{equmf}
\end{equation}

%%%%%%%%%%%%%%%%%%%%%%%%%%%%%%%%%%%%%%%%%%%%%%%%%%%%%%%%%%%%%%%%%%%%%
\subsection{Exact expressions for heat and work}
%%%%%%%%%%%%%%%%%%%%%%%%%%%%%%%%%%%%%%%%%%%%%%%%%%%%%%%%%%%%%%%%%%%%%

To solve the model exactly, we consider the limit in which the period $\tau$ is large compared to the relaxation times of the system. In this limit, 
at the end of the first and third parts of the cycle depicted in Fig.~\ref{fig1}, the system has reached an equilibrium stationary state. The probability 
distribution of spin configurations at the end of the first part of the cycle, denoted by $P^{(1)}_{\sigma}$, is then given by the Boltzmann distribution
with inverse temperature $\beta_c$ and magnetic field $H_1$. This can be written as
\begin{equation}
P^{(1)}_{\sigma} \propto \mathrm{e}^{-\beta_c E^{(1)}_\sigma},
\end{equation}
where $E^{(1)}_\sigma$ denotes the energy of configuration $\sigma$ with magnetic  field $H_1$. 
The same reasoning applies to the third part of the cycle, for which the probability distribution of spin configurations at the end of the third part is
\begin{equation}
P^{(3)}_{\sigma} \propto \mathrm{e}^{-\beta_h E^{(3)}_\sigma},
\end{equation}
where $E^{(3)}_\sigma$ is the energy with magnetic field $H_2$. We emphasize that, even in this large-$\tau$ limit, the protocol is not quasi-static, since the changes in temperature are assumed to be instantaneous.

In this limit, the average extracted work per cycle, $W$, and the average heat absorbed from the hot reservoir per cycle, $Q_h$, are given by the following expressions. 
The work extraction occurs during the instantaneous changes of the magnetic field at the end of the first and third parts of the cycle. If the system is in 
configuration $\sigma$ at the end of the first part, the energy change associated with switching the magnetic field from $H_1$ to $H_2$ is $E^{(3)}_\sigma - E^{(1)}_\sigma$. 
Similarly, at the end of the third part, the energy change associated with switching the field back from $H_2$ to $H_1$ is $E^{(1)}_\sigma - E^{(3)}_\sigma$. 
Since the work extracted from the system is equal to minus the change in its internal energy during these instantaneous steps, the average extracted work per cycle is given by
\begin{equation}
W = \frac{1}{N} \sum_\sigma \left(P^{(1)}_{\sigma} - P^{(3)}_{\sigma}\right)
\left(E^{(1)}_\sigma - E^{(3)}_\sigma\right).
\label{eqwork1}
\end{equation}
The factor $N^{-1}$ ensures that $W$ corresponds to the extracted work per spin. In the thermodynamic limit $N \to \infty$, both the work and heat per spin remain finite.

The average  heat extracted from the hot reservoir is the final average energy  minus the initial average energy in the third part of the cycle in Fig. \ref{fig1}, since the duration of the fourth part of the cycle is negligible. Hence,
the average heat (per spin) extracted from the hot reservoir per period is   
\begin{equation}
Q_h= N^{-1}\sum_\sigma(P^{(3)}_{\sigma}-P^{(1)}_{\sigma})E^{(3)}_\sigma.
\label{eqheat1}
\end{equation}
This expression is valid for the present protocol with instantaneous temperature changes.
The efficiency of the heat engine defined as
\begin{equation}
\eta \equiv \frac{W}{Q_h} \leq \eta_C \equiv 1 - \frac{\beta_h}{\beta_c},
\end{equation}
where $\eta_C$ is the Carnot efficiency.

Using the definitions of the average magnetization per spin in Eq.~\eqref{eqm} and the average energy per spin in Eq.~\eqref{equ}, the work expression in Eq.~\eqref{eqwork1} can be rewritten as
\begin{equation}
W = (H_2 - H_1)\left(m_{\beta_c,H_1,J} - m_{\beta_h,H_2,J}\right).
\label{eqwork2}
\end{equation}
Similarly, the heat absorbed from the hot reservoir in Eq.~\eqref{eqheat1} can be expressed as
\begin{equation}
Q_h = u_{\beta_h,H_2,J} - u_{\beta_c,H_1,J} + (H_2 - H_1)\, m_{\beta_c,H_1,J}.
\end{equation}

In the following, we focus on the average work per cycle rather than the power, which is defined as the work per cycle divided by the period $\tau$. 
In the large-$\tau$ limit considered here, the power is proportional to the average work per cycle, since increasing $\tau$ beyond the value required 
for equilibration only leads to a reduction of the power. In Sec.~\ref{sec4}, we perform numerical simulations for finite values of $\tau$ and analyze 
the dependence of the power on the period $\tau$.

%%%%%%%%%%%%%%%%%%%%%%%%%%%%%%%%%%%%%%%%%%%%%%%%%%%%%%%%%%%%%%%%%%%%%
\subsection{Model without interactions}
%%%%%%%%%%%%%%%%%%%%%%%%%%%%%%%%%%%%%%%%%%%%%%%%%%%%%%%%%%%%%%%%%%%%%

We first comment on the phenomenology of the model in the absence of interactions, corresponding to $J=0$, before turning to our investigation of the role of interactions. 
For $J=0$, the dimensionality of the model is irrelevant. In fact, even a system consisting of a single spin yields the same expression for the work per spin. 
This single-spin model is similar to a two-state system that has been analyzed previously in Ref.~\cite{espo10}.

The expression for the work can be obtained from Eq.~\eqref{eqwork2} by setting $J=0$ and using the corresponding magnetization from Eq.~\eqref{eqm1d}. This procedure yields
\begin{equation}
W_0 = (H_2 - H_1)\left(
\frac{\sinh(\beta_c H_1)}{\sqrt{\sinh^2(\beta_c H_1) + 1}}-\frac{\sinh(\beta_h H_2)}{\sqrt{\sinh^2(\beta_h H_2) + 1}}
\right),
\end{equation}
where $W_0$ denotes the extracted work per spin in the noninteracting case ($J=0$).

For the system to operate as a heat engine, corresponding to $W_0 > 0$, the parameters must satisfy the conditions 
$H_2 > H_1$ and $\beta_c H_1 > \beta_h H_2$. The efficiency in the absence of interactions can be obtained by 
computing the absorbed heat $Q_h$ for $J=0$. In this case, the efficiency takes the simple form
\begin{equation}
\eta = 1 - \frac{H_1}{H_2}.
\end{equation}
This efficiency combined with the condition $\beta_c H_1 > \beta_h H_2$ shows that the standard thermodynamic bound $\eta \leq \eta_C$ is fulfilled.

%%%%%%%%%%%%%%%%%%%%%%%%%%%%%%%%%%%%%%%%%%%%%%%%%%%%%%%%%%%%%%%%%%%%%
\section{Results}
%%%%%%%%%%%%%%%%%%%%%%%%%%%%%%%%%%%%%%%%%%%%%%%%%%%%%%%%%%%%%%%%%%%%%
\label{sec3}
%%%%%%%%%%%%%%%%%%%%%%%%%%%%%%%%%%%%%%%%%%%%%%%%%%%%%%%%%%%%%%%%%%%%%
\subsection{Heat engine with the 1D Ising model}
%%%%%%%%%%%%%%%%%%%%%%%%%%%%%%%%%%%%%%%%%%%%%%%%%%%%%%%%%%%%%%%%%%%%%

We first analyze the 1D model, which allows for a fully exact solution without the need to solve a transcendental equation. 
Using the expression for the magnetization in Eq.~\eqref{eqm1d} and the work in Eq.~\eqref{eqwork2}, the condition for the system to operate as a heat engine, $W>0$, can be written as
\begin{equation}
\log\!\left(\frac{\sinh(\beta_c H_1)}{\sinh(\beta_h H_2)}\right) > -2J(\beta_c-\beta_h).
\label{eqresult1}
\end{equation}
This exact condition provides a clear illustration of a central result of our work: interactions can transform a protocol, specified here by the inverse temperatures and magnetic fields, that would otherwise not produce work into a functioning heat-engine cycle. In the absence of interactions ($J=0$), the condition for engine operation reduces to $\beta_c H_1 > \beta_h H_2$. For nonzero interaction strength $J$, this restriction is relaxed, and the system can function as a heat engine even when $\beta_c H_1 \le \beta_h H_2$, provided that Eq.~\eqref{eqresult1} is satisfied. In all results presented in this section and the next, we fix $\beta_c=1$ and vary $0 \le \beta_h \le 1$.

\begin{figure}
\centering
\includegraphics{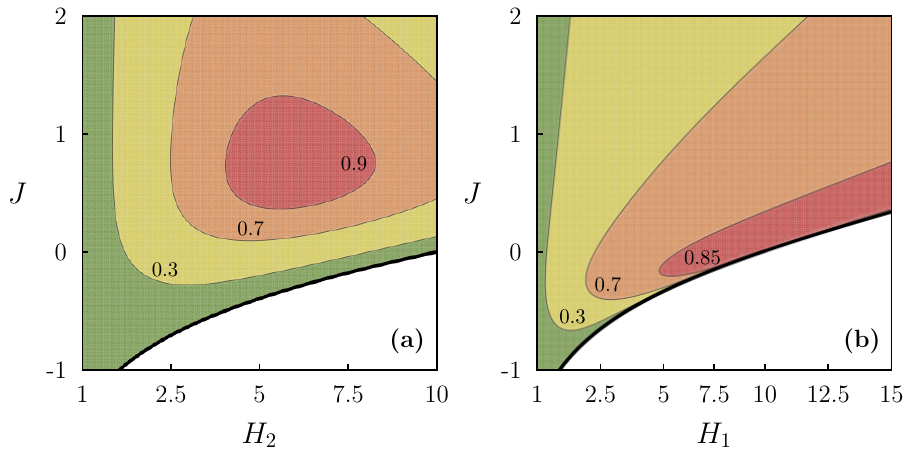}
\vspace{-2mm}
\caption{Contour plots for the 1D model in the $H_2 \times J$ plane. (a) Work $W$. (b) Efficiency $\eta$. The remaining parameters are set to $\beta_h=0.1$ and $H_1=1$.}
\label{fig2}
\end{figure}

Figure~\ref{fig2} shows contour plots of the work $W$ and the efficiency $\eta$ in the $H_2 \times J$ plane. For a fixed protocol, i.e., fixed values of $\beta_c$, $\beta_h$, $H_1$, and $H_2$, 
interactions quantified by the parameter $J$ can enhance both the work output and the efficiency. 
In all cases observed, the work is maximized for ferromagnetic interactions ($J>0$). The efficiency, however, 
can be maximized for either positive or negative values of $J$, depending on the remaining parameters. 
This behavior can be understood as follows. 
Consider the choice $H_1=\beta_c^{-1}=1$, as in Fig.~\ref{fig2}. If $H_2=\beta_h^{-1}$, the efficiency approaches the Carnot efficiency for $J=0$, which therefore corresponds to the optimal efficiency. 
If $H_2<\beta_h^{-1}$, the efficiency is maximized for $J<0$, whereas for $H_2>\beta_h^{-1}$ it is maximized for $J>0$.

We now turn to the optimization of the work, or equivalently the power, since both are proportional in the large-$\tau$ limit, with respect to the external fields $H_1$, $H_2$, and the interaction strength $J$, 
for a fixed value of $\beta_h \le 1$. We denote the corresponding maximum work by $W^*$, and the associated efficiency—referred to as the efficiency at maximum power—by $\eta^*$. 
These results are compared to the maximum work obtained for $J=0$, optimized with respect to $H_1$ and $H_2$ at fixed $\beta_h$. 
The maximum work in the noninteracting case is denoted by $W_0^*$, and the corresponding efficiency at maximum power by $\eta_0^*$. 
We also compare these efficiencies to the Curzon--Ahlborn efficiency \cite{curz75},
\begin{equation}
\eta_{\mathrm{CA}} \equiv 1 - \sqrt{\beta_h/\beta_c},
\end{equation}
which is a well-known expression for the efficiency at maximum power obtained within linear response theory and beyond it \cite{espo09}.

\begin{figure}
\centering
\includegraphics{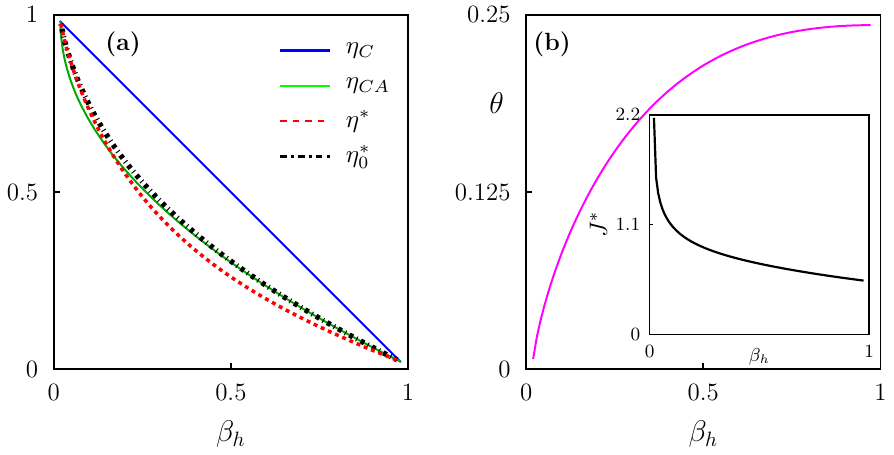}
\vspace{-2mm}
\caption{Efficiency at maximum power for the 1D model. (a) Efficiency at maximum power $\eta^*$ as a function of $\beta_h$. (b) Ratio $\theta$ as a function of $\beta_h$.}
\label{fig3}
\end{figure}

Our results for the efficiency at maximum power are shown in Fig.~\ref{fig3}. The curve corresponding to $\eta_0^*$ lies close to the Curzon--Ahlborn efficiency $\eta_{\mathrm{CA}}$, 
whereas the efficiency $\eta^*$ that incorporates interactions remains below $\eta_0^*$. One possible explanation is that, 
for $J=0$, there is tight coupling between heat and work \cite{espo09}, leading to a simple expression for the efficiency, $\eta_0 = 1 - H_1/H_2$. 
In contrast, for $J \neq 0$ the efficiency acquires a more involved dependence on the interaction strength. 

Despite the fact that $\eta_0^* > \eta^*$, the maximum work per cycle in the presence of interactions exceeds the maximum work obtained without interactions. 
We quantify this enhancement through the ratio
\begin{equation}
\theta \equiv \frac{W^* - W_0^*}{W_0^*},
\label{eqtheta}
\end{equation}
which is shown in Fig.~\ref{fig3}(b). In the inset of the right panel of Fig.~\ref{fig3}, we also display the optimal interaction strength 
$J^*$ that maximizes the work. We find that $J^*>0$ for all values of $\beta_h$, indicating that the maximum work is always achieved
 for ferromagnetic interactions.

%%%%%%%%%%%%%%%%%%%%%%%%%%%%%%%%%%%%%%%%%%%%%%%%%%%%%%%%%%%%%%%%%%%%%
\subsection{Heat engine with the MF Ising model}
%%%%%%%%%%%%%%%%%%%%%%%%%%%%%%%%%%%%%%%%%%%%%%%%%%%%%%%%%%%%%%%%%%%%%

\begin{figure}
\centering%
 \includegraphics{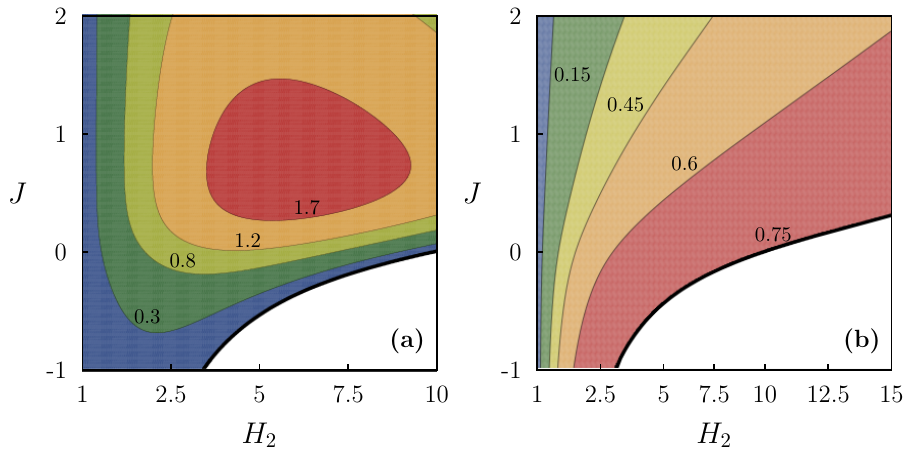}%
\vspace{-2mm}
\caption{Contour plots for the MF model in the $H_2\times J$ plane. (a) Work $W$. (b) Efficiency $\eta$. The other parameters are set to $\beta_h=0.1$ and $H_1=1$.}   
\label{fig4}
\end{figure}

For the MF model, we likewise observe that interactions can enhance both the work output and the efficiency of the engine, as shown in Fig.~\ref{fig4}. For fixed values of $\beta_c$, $\beta_h$, $H_1$, and $H_2$, the work is maximized for positive values of $J$, corresponding to ferromagnetic interactions. The efficiency, however, can be maximized for either positive or negative values of $J$, depending on the specific choice of these four parameters. Qualitatively, the behavior observed in this regime is similar to that found for the 1D model in Fig.~\ref{fig2}.

A genuinely new regime emerges in the MF model as a consequence of the phase transition, which allows for a nonzero magnetization even in the absence of an external magnetic field.  Setting $H_1=0$ in Eq.~\eqref{eqwork2}, the work reduces to
\begin{equation}
W = H_2 \left( m_{\beta_c,J} - m_{\beta_h,H_2,J} \right),
\label{eqworkreg1}
\end{equation}
where $m_{\beta_c,J}$ denotes the spontaneous magnetization defined in Eq.~\eqref{eqm0}. Clearly, from this expression we see that, without magnetization at zero field, it is impossible to obtain positive work when $H_1=0$.
The system can operate as a heat engine provided that $\beta_c$ and $J$ are such that the cold temperature lies below the critical temperature. Moreover, the spontaneous magnetization during the cold part 
of the cycle must exceed the magnetization induced by the external field $H_2$ during the hot part of the cycle.

A remarkable result emerges when analyzing the maximum work produced by this heat engine. We find that the maximum work $W^*$ is achieved precisely when the engine operates in this regime associated with spontaneous magnetization. Specifically, if $\beta_h$ is fixed and the work is maximized with respect to $H_1$, $H_2$, and $J$, the optimal point is numerically consistent with $H_1=0$ for all values of $\beta_h$ considered. This shows that the maximum power output of the MF heat engine is attained in a regime that relies on the phase transition.

\begin{figure}
\centering
\includegraphics[width=\textwidth]{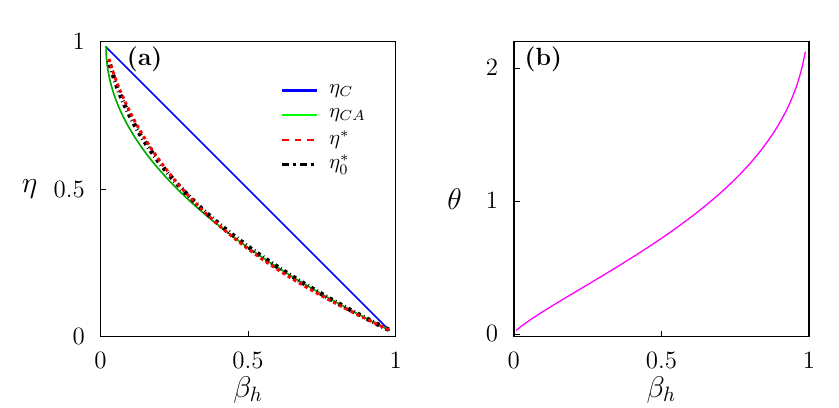}
\vspace{-2mm}
\caption{Efficiency at maximum power for the MF model. (a) Efficiency at maximum power $\eta^*$ as a function of $\beta_h$. (b) Ratio $\theta$ as a function of $\beta_h$.}
\label{fig5}
\end{figure}

In Fig.~\ref{fig5}(a), we show the efficiency at maximum power $\eta^*$ for the MF model. This curve lies very close to the corresponding efficiency at maximum power in the absence of interactions, $\eta_0^*$. 
We further compare the maximum work obtained with interactions to that obtained without interactions using the ratio $\theta$ defined in Eq.~\eqref{eqtheta}, shown in Fig.~\ref{fig5}(b). 
While the efficiency at maximum power follows a very similar trend in both cases, the maximum work in the presence of interactions can be significantly larger.

%%%%%%%%%%%%%%%%%%%%%%%%%%%%%%%%%%%%%%%%%%%%%%%%%%%%%%%%%%%%%%%%%%%%%
\subsection{Heat engine with the variation of the interaction parameter $J$}
%%%%%%%%%%%%%%%%%%%%%%%%%%%%%%%%%%%%%%%%%%%%%%%%%%%%%%%%%%%%%%%%%%%%%

For the MF model, we consider a different thermodynamic cycle from the one shown in Fig.~\ref{fig1}, which explicitly relies
 on the existence of a phase transition. In this cycle, the external magnetic field is kept fixed at $H=0$ throughout, 
 while the interaction strength takes the value $J_1$ during the cold part of the cycle and $J_2$ during the hot part. 
 This heat engine is only possible due to the presence of a phase transition that allows for spontaneous magnetization 
 at zero field. The expressions for work and heat differ from those obtained for the cycle considered previously. 
 This type of cycle with the variation of the interaction strength has been considered in \cite{camp16}, where 
 in their case the system is at the critical temperature throughout the cycle and the engine  can be arbitrarily close to 
 Carnot efficiency at finite power. Their results rely on finite-size scaling close to criticality, different for 
 the thermodynamic limit considered here.

Starting from Eq.~\eqref{eqwork1}, the extracted work per cycle for this regime is given by
\begin{equation}
W = \frac{J_2 - J_1}{2}\left(m^2_{\beta_c,J_1} - m^2_{\beta_h,J_2}\right),
\label{eqworkreg2}
\end{equation}
where $m_{\beta_c,J_1}$ denotes the spontaneous magnetization, defined as the solution of the transcendental equation in Eq.~\eqref{eqmmf} at inverse temperature 
$\beta_c$. From Eq.~\eqref{eqheat1}, the heat absorbed from the hot reservoir reads
\begin{equation}
Q_h = \frac{J_2}{2}\left(m^2_{\beta_c,J_1} - m^2_{\beta_h,J_2}\right).
\label{eqheatreg2}
\end{equation}
The efficiency of this heat engine is, therefore,
\begin{equation}
\eta \equiv \frac{W}{Q_h} = 1 - \frac{J_1}{J_2}.
\end{equation}

\begin{figure}
\centering
\includegraphics{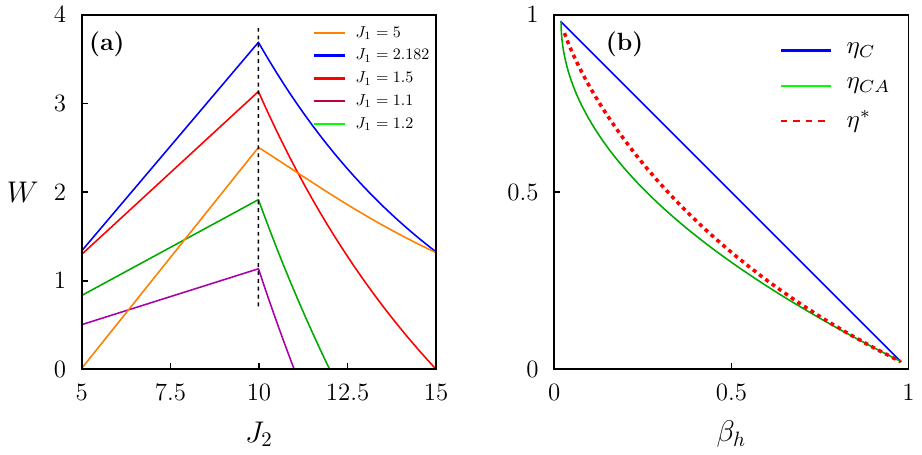}
\vspace{-2mm}
\caption{Results for the cycle with varying interaction strength $J$. (a) Work as a function of $J_2$ for fixed $J_1$ and $\beta_h=0.1$. (b) Efficiency at maximum power $\eta^*$.}
\label{fig6}
\end{figure}

For fixed inverse temperatures $\beta_c$ and $\beta_h$, the behavior of this heat engine can be summarized as follows. First, the parameter $J_1$ must satisfy $J_1 > \beta_c^{-1} = 1$
 in order for spontaneous magnetization to exist during the cold part of the cycle, i.e., $m_{\beta_c,J_1} > 0$. 
 Otherwise, no work can be extracted and $W=0$. Regarding the parameter $J_2$, extracted work is obtained even when $J_2 < \beta_h^{-1}$, 
 a regime in which the magnetization vanishes, $m_{\beta_h,J_2} = 0$. For $J_2 \le \beta_h^{-1}$, therefore, the work is  an increasing function of $J_2$, 
 since the contribution from the magnetization during the hot part of the cycle remains zero in this range.
When $J_2$ is increased beyond the critical value $\beta_h^{-1}$, the magnetization $m_{\beta_h,J_2}$ becomes nonzero and starts to contribute to the work. 
Interestingly, there is local maximum of work for any fixed value of $J_1$ at the critical point $J_2 = \beta_h^{-1}$. This behavior is illustrated in Fig.~\ref{fig6}.

The maximum work $W^*$ with respect to $J_1$ and $J_2$ is obtained in this regime with $J_2 = \beta_h^{-1}$. 
The corresponding efficiency at maximum power $\eta^*$ is shown in Fig.~\ref{fig6}. 
Remarkably, this efficiency curve lies significantly above the Curzon--Ahlborn efficiency $\eta_{\mathrm{CA}}$ 
over the entire range of $\beta_h$.

%%%%%%%%%%%%%%%%%%%%%%%%%%%%%%%%%%%%%%%%%%%%%%%%%%%%%%%%%%%%%%%%%%%%%
\section{Finite Period $\tau$}
%%%%%%%%%%%%%%%%%%%%%%%%%%%%%%%%%%%%%%%%%%%%%%%%%%%%%%%%%%%%%%%%%%%%%
\label{sec4}

\subsection{Model and results for power}

In this section we consider the MF model for a finite cycle period $\tau$. 
The objectives are to analyze the power 
\begin{equation}
P \equiv \frac{W}{\tau},
\end{equation}
as a function of $\tau$ for periods that are not necessarily long enough for the system to reach a stationary state and to verify the correctness of our analytical results for large $\tau$.

For the cycle shown in Fig.~\ref{fig1}, with instantaneous changes in temperature and a finite period $\tau$, the inverse temperature and the magnetic field can be written as explicit functions of time,
\begin{eqnarray}
\beta(t) =
\left\{
\begin{array}{ll}
\beta_c, & \text{if } 0 \le t < \tau/2, \\
\beta_h, & \text{if } \tau/2 \le t < \tau,
\end{array}
\right.
\end{eqnarray}
and
\begin{eqnarray}
H(t) =
\left\{
\begin{array}{ll}
H_1, & \text{if } 0 \le t < \tau/2, \\
H_2, & \text{if } \tau/2 \le t < \tau.
\end{array}
\right.
\end{eqnarray}
For the standard heat-engine cycle, the interaction strength $J$ is kept constant. For the MF Ising heat engine with vanishing magnetic field, $H(t)=0$, the interaction strength is instead time-dependent and given by
\begin{eqnarray}
J(t) =
\left\{
\begin{array}{ll}
J_1, & \text{if } 0 \le t < \tau/2, \\
J_2, & \text{if } \tau/2 \le t < \tau.
\end{array}
\right.
\end{eqnarray}

For numerical simulations of the MF model at finite period, it is convenient to use the total magnetization $M_\sigma = \sum_{i=1}^N \sigma_i$ as a coarse-grained variable. The magnetization $M$ takes the values
$M = -N, -N+2, \ldots, N-2, N$, where $N$ is assumed to be even. A single spin flip corresponds to a step of size two in this magnetization space.
The number of microscopic spin configurations corresponding to a given magnetization $M$ is
$N!/(N_+! N_-!)$, where $N_\pm = (N \pm M)/2$. Furthermore, the MF Ising energy in Eq.~\eqref{eqEmf}, written in terms of the magnetization $M$ and keeping the explicit time dependence of the parameters, reads
\begin{equation}
E_M(t) = -\frac{J(t)}{2N} M^2 - H(t) M.
\label{eqEsim}
\end{equation}

\begin{figure}
\centering
\includegraphics[width=\textwidth]{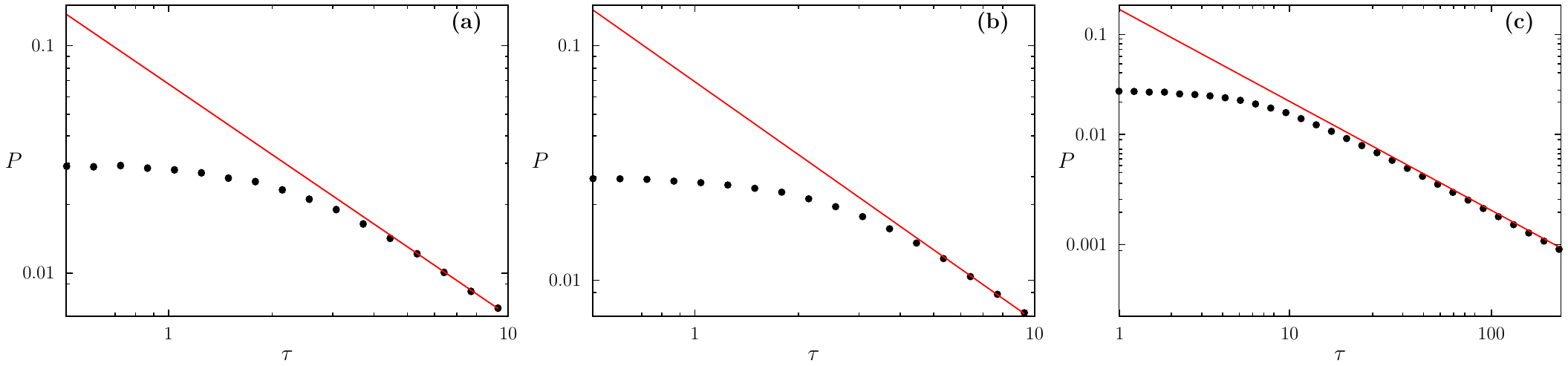}
\vspace{-2mm}
\caption{Power $P$ as a function of the period $\tau$ obtained from numerical simulations. The inverse hot temperature is $\beta_h=0.5$. The kinetic rate is $k=1$, and the time step is $\delta t = 0.01/(2N)$. In all panels, the red solid line corresponds to $W/\tau$, where the work $W$ is computed from the analytical expressions. Agreement with numerical results is observed for sufficiently large $\tau$. (a) $H_1=1$, $H_2=2$, $J=1$, $N=200$. (b) $H_1=0$, $H_2=1$, $J=2$, $N=200$. (c) Engine with $H=0$ and varying interaction strength: $J_1=1.5$, $J_2=2$, $N=800$.}
\label{fig7}
\end{figure}

The transition rates for this random walk in magnetization space are time-periodic and read as follows. A transition from $M$ to $M+2$ occurs with rate
\begin{equation}
k_{M \to M+2}(t) \equiv
\frac{k (N - M)\, \mathrm{e}^{\beta(t)\left[J(t)\left(m + N^{-1}\right) + H(t)\right]}}
{2 \cosh\!\left\{\beta(t)\left[J(t)\left(m + N^{-1}\right) + H(t)\right]\right\}},
\label{eqrates1}
\end{equation}
where $k$ is a kinetic parameter that sets the overall time scale of the transitions and $m\equiv M/N$.
The transition rate from $M$ to $M-2$ is given by
\begin{equation}
k_{M \to M-2}(t) \equiv
\frac{k (N + M)\, \mathrm{e}^{-\beta(t)\left[J(t)\left(m - N^{-1}\right) + H(t)\right]}}
{2 \cosh\!\left\{\beta(t)\left[J(t)\left(m - N^{-1}\right) + H(t)\right]\right\}}.
\label{eqrates2}
\end{equation}
These rates satisfy detailed balance with respect to the energy function in Eq.~\eqref{eqEsim}. For the numerical implementation, 
we discretize time using a small time step $\delta t$ such that further reductions in $\delta t$ do not lead to appreciable changes in the results.

The numerical results for the power $P$ as a function of the period $\tau$ are presented in Fig.~\ref{fig7} for three different cases: both magnetic fields nonzero, 
one magnetic field set to zero, and the cycle with time-dependent interaction strength. For all cases considered, the power is found to be a decreasing function of $\tau$. 
Due to the absence of analytical results at finite $\tau$, we cannot claim that this monotonic behavior holds for all parameter values. 
On a different note, the numerical results confirm the correctness of the analytical calculations in the long-period limit,  
as indicated by the solid red lines in Fig.~\ref{fig7}.

\subsection{Changes between positive and negative spontaneous magnetization}

Our numerical results also serve to confirm the validity of the analytical calculations in the long-period limit, as illustrated by the solid red lines in Fig.~\ref{fig7}. We now turn to the second and third cases in Fig.~\ref{fig7}, focusing on the possibility of the system switching between the two values of the spontaneous magnetization.

First, consider the case in which one of the magnetic fields is set to zero, corresponding to panel (b) in Fig.~\ref{fig7}. During the first stage of the period 
(it corresponds to the third part of the cycle in Fig.~\ref{fig1}, but the starting point of a cyclic protocol is arbitrary), the external field is positive, and 
the system relaxes to a unique state with positive magnetization. When the system enters the second stage, it retains a positive magnetization and therefore preferentially relaxes toward the positive branch of the spontaneous magnetization. 
Importantly, this mechanism does not rely on proximity to the critical point. Rather, for the work to be positive, the spontaneous magnetization in the second stage must exceed the magnetization induced by the external field in the first stage. Transitions between the two branches occur on timescales that grow exponentially with system size. Accordingly, we consider periods $\tau$ that are long compared to the relaxation time toward the positive spontaneous magnetization, given a positive initial magnetization, but short compared to the exponentially large time required to overcome the energy barrier separating the two branches. For moderate system sizes, this latter timescale is already prohibitively large.

 Second, consider the cycle with time-dependent interaction strength. In this case, 
 the system can operate close to the critical point, and branch switching becomes possible. 
 Specifically, $J_1$ must exceed the critical value $\beta_c^{-1}$ so that spontaneous 
 magnetization arises during the first stage of the cycle. In the second stage, one may have $J_2 \le \beta_h^{-1}$, 
 in which case no spontaneous magnetization exists.
As a result, when the system enters the stage with spontaneous magnetization, 
it effectively starts from an initial condition with vanishing average magnetization. 
Consequently, for $J_2 \le \beta_h^{-1}$, the system can switch between the two branches of the spontaneous 
magnetization across different cycles. However, such branch switching does not affect the extracted work, for the cycle 
with time-dependent interaction strength, the work in Eq.~\eqref{eqworkreg2} depends quadratically on the magnetization.

 In summary, when one of the magnetic fields is set to zero, the large energy barrier suppresses branch switching, which makes our analytical prediction that uses the 
 positive spontaneous magnetization consistent with the numerical results. For the protocol with time-dependent interaction strength, branch switching can occur but it does not affect 
 the extracted work. For these reasons, hysteresis \cite{kund23,wu25,chen26,chen26b} does not play a role in our engine. However, hysteresis could become relevant for protocols with a different time dependence 
 of the control parameters, where it may influence both the power and the efficiency.

%%%%%%%%%%%%%%%%%%%%%%%%%%%%%%%%%%%%%%%%%%%%%%%%%%%%%%%%%%%%%%%%%%%%%
\section{Conclusion}
%%%%%%%%%%%%%%%%%%%%%%%%%%%%%%%%%%%%%%%%%%%%%%%%%%%%%%%%%%%%%%%%%%%%%
\label{sec5}

We have analyzed the role of interactions and phase transitions in the performance of a cyclic heat engine whose system is the Ising model. 
We have shown that interactions can enhance both the work output and the efficiency. Moreover, interactions can transform a protocol that would 
not lead to work extraction in the absence of interactions into a functional heat engine. This behavior is demonstrated explicitly for the 1D model through 
a simple exact expression in Eq. \eqref{eqresult1}.

The MF model provides the additional feature of a phase transition, which allows us to investigate the role of criticality in the performance of the engine. In particular, 
the presence of the phase transition makes it possible for the system to operate as a heat engine even when one of the magnetic fields is set to zero. 
This regime is only accessible due to spontaneous magnetization, the work given in Eq. \eqref{eqwork2} cannot be positive for $H_1=0$ if the corresponding magnetization is also zero. 
Remarkably, for a fixed temperature difference, the work output is maximized precisely in this regime with one magnetic field set to zero. Hence,
the phase transition plays a central role in the optimization of work output.

We have also analyzed a cycle in which the interaction strength $J$ varies while the magnetic field is kept at zero throughout the entire cycle. 
This engine likewise relies on spontaneous magnetization for its operation. Interestingly, the efficiency at maximum power for this cycle 
remains above the well-known Curzon-Ahlborn efficiency, which is valid within linear response theory.

Furthermore, we performed numerical simulations of the MF model for finite periods, which lead to two main observations. First, the power output is a monotonically 
decreasing function of the period; we did not observe any regime in which an intermediate period yields optimal power. This result is limited to our numerical analysis
 and has not yet been confirmed by an analytical study at finite period. Second, for sufficiently long periods, the numerical results agree with our analytical predictions, 
 providing a consistency check of our calculations.

We expect that the enhancement of work (or power) and efficiency due to interactions is a generic feature that should also occur in other interacting systems. 
Whether the optimal performance of the engine is generically achieved in a regime that exploits a phase transition, as found here for the MF model, remains an open question. 
Investigating the generality of this feature across different models constitutes an interesting direction for future research. Furthermore, the analysis of power 
and efficiency for protocols with more complex time-dependent parameters, which may exhibit phenomena such as hysteresis, remains an open problem.
Finally, we note that our setup provides an appealing pedagogical example of a cyclic heat engine, as the heat and work are expressed in terms of standard textbook 
formulas for the magnetization of the 1D and MF Ising models.

\ack
A . C. B. acknowledges financial support from the NSF through the grant  DMR-2424140. G. A. L. F. and 
C. E. F. acknowledge the financial support from FAPESP under Grants 2022/15453-0, 2022/16192-5 and 2024/03763-0. The financial support from CNPq is also acknowledged.

%==========================================================================
% References
%==========================================================================

\section*{References}

\bibliography{refs} 

% \bibliographystyle{unsrt}
% \input{thermocellpaper.bbl}

% \bibliography{thermocellpaper.bbl} 

 \end{document}